\documentclass[conference]{IEEEtran}
\ifCLASSINFOpdf
   \usepackage[pdftex]{graphicx}
\else
\fi
\usepackage{algorithmic}
\usepackage{algorithm}

\usepackage{amsmath,mathtools}
\usepackage{amssymb}
\usepackage{amsfonts}
\usepackage{graphicx}
\usepackage{amsbsy}
\usepackage[margin = 10pt]{caption}
\usepackage[caption=false,font=footnotesize]{subfig}

\usepackage[top=0.71in, left = 0.67in, right = 0.67in, bottom = 0.99in]{geometry}
\newcommand {\vc} [1] {\boldsymbol{#1}}
\begin{document}
%
\title{How Can Subsampling Reduce Complexity in Sequential MCMC Methods and Deal with Big Data in Target Tracking?}



\author{\IEEEauthorblockN{Allan De Freitas$^*$, Fran\c{c}ois Septier$^{\S}$, Lyudmila Mihaylova$^*$, Simon Godsill$^{ \natural}$}
\IEEEauthorblockA{$^*$Department of Automatic Control and Systems Engineering, University of Sheffield, United Kingdom\\
$^{\S}$ Institute Mines Telecom/Telecom Lille, CRIStAL UMR CNRS 9189, France\\
$^{ \natural}$Department of Engineering, Cambridge University, CB
1PZ, United
Kingdom\\
 Emails: \emph{a.defreitas@sheffield.ac.uk, francois.septier@telecom-lille.fr, l.s.mihaylova@sheffield.ac.uk, sjg@eng.cam.ac.uk} }}


%


\maketitle

\begin{abstract}
Target tracking faces the challenge in coping with large volumes of
data which requires efficient methods for real time applications.
The complexity considered in this paper is when there is a large
number of measurements which are required to be processed at each
time step. Sequential Markov chain Monte Carlo (MCMC) has been shown
to be a promising approach to target tracking in complex
environments, especially when dealing with clutter. However, a large
number of measurements usually results in large processing
requirements. This paper goes beyond the current state-of-the-art
and presents a novel Sequential MCMC approach that can overcome this
challenge through adaptively subsampling the set of measurements.
Instead of using the whole large volume of available data, the
proposed algorithm performs a trade off between the number of
measurements to be used and the desired accuracy of the estimates to
be obtained in the presence of clutter. We show results with large
improvements in processing time, more than 40 \% with a negligible
loss in tracking performance, compared with the solution without
subsampling.
\end{abstract}


%
\IEEEpeerreviewmaketitle

\section{Introduction}
Flooded with data, richly provided by electronic sensors, the
current monitoring systems face the problem of being able to process
the data and monitor the phenomenon of interest at the same time.
%
In this paper we consider the problem of target tracking in large
volumes of data.
%
%
There is a wealth of algorithms that can provide sequential
estimation of the states of the target, e.g. for details
see~\cite{Mihaylova2014,Balakrishnan2006}. In a Bayesian framework, the posterior
distribution can be iteratively computed. However,
analytically this can be achieved only when the state space model is
linear and perturbed by a Gaussian noise. In this case the solution
is referred to as the Kalman Filter. There are a large number of
techniques which overcome the limitations of the Kalman filter based
on the sequential Monte Carlo (SMC) methodology. The seminal work on SMC in target
tracking was presented in \cite{Gordon1993} which was also referred
to as the bootstrap particle filter (PF). The bootstrap PF and many
variants thereof, broadly referred to as PFs, are commonly favoured
techniques in a wide variety of applications due to the filters
ability to handle non-linear state space models and/or state space
models perturbed by non-Gaussian noise. However, the PF is not void
of challenges. Some of the difficulties faced by PFs includes weight
degeneracy and sample impoverishment. Although there are variants of
the PF which have been proposed to alleviate these issues
\cite{Gilks2001,Djuric2013}, the PF is still susceptible to degeneracy, and
these difficulties are more profound when tracking complex
systems.

Markov chain Monte Carlo (MCMC) techniques are a powerful set of algorithms for sampling from a probability distribution. MCMC tecnhiques, such as the Metropolis Hastings (MH) algorithm, have been predominantly used in applications requiring static inference \cite{Jasra2007}.
Recently there has been considerable interest in extending these techniques to sequentially updating the posterior distribution \cite{Khan2005,Septier2009}. Sequential MCMC has shown promising results for complex systems. The largest hindrance being long processing times which could limit usage in applications required to run in real time. There have also been several algorithms \cite{Korattikara2014, Bardenet2014, Bardenet2015a} which have been proposed to help reduce computational complexities when performing static inference with MCMC techniques on large datasets. 

In this paper we propose a novel technique which results in an efficient sequential MCMC algorithm when applied in complex systems consisting of a large number of measurements. This is achieved through the combination of sequential inference and adaptive subsampling of the measurements at each time step. We show how the proposed adaptive subsampling sequential MCMC algorithm can be applied to target tracking and illustrate the computational savings it affords. 
\section{Problem Formulation}
Target tracking of a complex system can be considered as sequential state estimation with multiple measurements. This can be achieved in a Bayesian framework by sequentially computing the filtering posterior distribution $p(\vc{x}_k|\vc{z}_{1:k})$ where $\vc{x}_k \in \mathbb{R}^{n_{\vc{x}}}$ is the state vector at time $t_k$ with $k = {1,...,T} \in \mathbb{N}$, and $\vc{z}_{1:k} = \{\vc{z}_1,...,\vc{z}_k\}$, represents all the measurements received up till time $t_k$. The measurements received at each time $t_k$ are represented by a set $\vc{z}_k = \{\vc{z}_k^1,...,\vc{z}_k^{M_k}\}$, where $M_k$ is the total number of measurements and $\vc{z}_k^i \in \mathbb{R}^{n_{\vc{z}}}$. The filtering posterior distribution can be recursively updated based on
\begin{equation}
p(\vc{x}_k|\vc{z}_{1:k}) \propto \int p(\vc{z}_k|\vc{x}_k) p(\vc{x}_k|\vc{x}_{k-1})p(\vc{x}_{k-1}|\vc{z}_{1:k-1})d\vc{x}_{k-1},\label{filterpdf}
\end{equation}
where $p(\vc{z}_k|\vc{x}_k)$ is referred to as the likelihood probability density function (pdf), and $p(\vc{x}_k|\vc{x}_{k-1})$ is referred to as the state transition pdf. An analytical solution to \eqref{filterpdf} is typically intractable when the state space model is characterised by non-linearities and/or non-Gaussian noise.
\subsection{Sequential Markov Chain Monte Carlo}
MCMC methods work by constructing a Markov chain with a desired distribution as the equilibrium distribution. A common MCMC technique used to obtain samples from the equilibrium distribution, $\pi(\vc{x})$, is the MH algorithm. This is achieved by first generating a sample from a known proposal distribution $\vc{x}^*\sim q(\,\cdot\,|\vc{x}^{m-1})$. The proposed sample is accepted as the current state of the chain, $\vc{x}^m$, if the following condition is satisfied
\begin{equation}
u < \frac{\pi(\vc{x}^*)q(\vc{x}^{m-1}|\vc{x}^*)}{\pi(\vc{x}^{m-1})q(\vc{x}^*|\vc{x}^{m-1})},
\end{equation}
where $u$ represents a sample from a uniform random variable $u \sim \textit{U}_{[0,1]}$. Using Bayes' rule and assuming that there are $M$ conditionally independent measurements, $\vc{z}^i$, results in the further expansion of this expression
\begin{equation}
u < \frac{p(\vc{x}^*)q(\vc{x}^{m-1}|\vc{x}^*)}{p(\vc{x}^{m-1})q(\vc{x}^*|\vc{x}^{m-1})}\prod_{i=1}^M \frac{p(\vc{z}^i|\vc{x}^*)}{p(\vc{z}^i|\vc{x}^{m-1})}.
\end{equation}
The previous state of the chain is stored as the current state, $\vc{x}^m = \vc{x}^{m-1}$, when the proposed sample does not meet this criterion. We further manipulate this expression into a form with the likelihood isolated:
\begin{equation}
\log\left[u \frac{p(\vc{x}^{m-1})q(\vc{x}^*|\vc{x}^{m-1})}{p(\vc{x}^*)q(\vc{x}^{m-1}|\vc{x}^*)}\right] <\sum_{i=1}^M \log\left[\frac{p(\vc{z}^i|\vc{x}^*)}{p(\vc{z}^i|\vc{x}^{m-1})}\right].
\end{equation}

In \cite{Khan2005} it was proposed to use MCMC methods, specifically the MH algorithm, to target the filtering posterior distribution in \eqref{filterpdf} as the equilibrium distribution. This allows for the iterative update of an approximation of the filtering posterior distribution by representing $p(\vc{x}_{k-1}|\vc{z}_{1:k-1})$ with a set of unweighted particles,
\begin{equation}
p(\vc{x}_{k-1}|\vc{z}_{1:k-1}) \approx \frac{1}{N_p}\sum_{j=1}^{N_p}\delta(\vc{x}_{k-1}-\vc{x}_{k-1}^{(j)}),
\end{equation}
 where $N_p$ is the number of particles and $(j)$ the particle index. This technique was shown to work well in state space models containing a high number of dimensions when compared to techniques relying on importance sampling, however, this direct approach may result in a high computational expense.

It was proposed in \cite{Septier2009} to consider targeting the joint filtering posterior distribution of $\vc{x}_k$ and $\vc{x}_{k-1}$
\begin{equation}
p(\vc{x}_k,\vc{x}_{k-1}|\vc{z}_{1:k}) \propto  p(\vc{z}_k|\vc{x}_k) p(\vc{x}_k|\vc{x}_{k-1})p(\vc{x}_{k-1}|\vc{z}_{1:k-1}),
 \end{equation} 
as the equilibrium distribution in order to help alleviate the high computational demand. In a similar fashion, an approximation for the joint filtering posterior distribution can be obtained through MCMC methods by representing $p(\vc{x}_{k-1}|\vc{z}_{1:k-1})$ with a set of unweighted particles. This approach has the advantage of avoiding the direct Monte Carlo computation of the predictive posterior density. Furthermore, the approximation can be marginalised to obtain the filtering posterior distribution of interest. 

More specifically, at each time step, the particles are updated with a MH joint draw for $\vc{x}_k$ and $\vc{x}_{k-1}$, followed by an individual MH draw for $\vc{x}_k$. The  second step, referred to as the refinement step, is introduced to aid in the mixing of the chain. An appropriate burn in period, $N_{burn}$, was also introduced to minimize the effect of the initial values of the Markov chain. This results in the definition of the total number of MCMC iterations at each time step, $N = N_p + N_{burn}$. This approach is highlighted by Algorithm~\ref{alg1} and is referred to as standard sequential MCMC. This approach showed promising results in a multi-target environment but is still susceptible to high computational complexity when a substantially large amount of measurements are required to be processed. 
 \begin{algorithm}[!ht]
\caption{Sequential Markov Chain Monte Carlo}
\label{alg1}
\begin{algorithmic}[1]
\STATE Initialize particle set: $\{\vc{x}_0^{(j)}\}_{j=1}^{N_p}$
\FOR{$k$ = 1,...,$T$}
\FOR{$m$ = 1,...,$N$}
\STATE \textit{\underline{Joint Draw}}
\STATE Propose $\{\vc{x}_k^*,\vc{x}_{k-1}^*\} \sim q_1\left(\vc{x}_k,\vc{x}_{k-1}|\vc{x}_k^{m-1},\vc{x}_{k-1}^{m-1}\right)$
\STATE Compute $\psi_1(u,\vc{x}_k^*,\vc{x}_{k-1}^*,\vc{x}_k^{m-1},\vc{x}_{k-1}^{m-1})$ \\$= \frac{1}{M_k}\log\biggl[u\frac{p(\vc{x}_k^{m-1}|\vc{x}_{k-1}^{m-1})p(\vc{x}_{k-1}^{m-1}|\vc{z}_{1:k-1})}{p(\vc{x}_k^*|\vc{x}_{k-1}^*)p(\vc{x}_{k-1}^*|\vc{z}_{1:k-1})}\times$\\ \hspace{40mm}$\frac{q_1\left(\vc{x}_k^*,\vc{x}_{k-1}^*|\vc{x}_k^{m-1},\vc{x}_{k-1}^{m-1}\right)}{q_1\left(\vc{x}_k^{m-1},\vc{x}_{k-1}^{m-1}|\vc{x}_k^*,\vc{x}_{k-1}^*\right)}\biggr]$
\STATE Compute $\Lambda_1^{M_k}(\vc{x}_k^*,\vc{x}_k^{m-1})$ \\ \hspace{25mm}$= \frac{1}{M_k}\sum_{i=1}^{M_k}\log\left[\frac{p(\vc{z}_k^i|\vc{x}_k^*)}{p(\vc{z}_k^i|\vc{x}_k^{m-1})}\right]$
\IF{{$\Lambda_1^{M_k}(\vc{x}_k^*,\vc{x}_k^{m-1})$ \\ \hspace{15mm}$>\psi_1(u,\vc{x}_k^*,\vc{x}_{k-1}^*,\vc{x}_k^{m-1},\vc{x}_{k-1}^{m-1})$}}
\STATE $\{\vc{x}_k^m,\vc{x}_{k-1}^m\} = \{\vc{x}_k^*,\vc{x}_{k-1}^*\}$
\ELSE
\STATE $\{\vc{x}_k^m,\vc{x}_{k-1}^m\} = \{\vc{x}_k^{m-1},\vc{x}_{k-1}^{m-1}\}$
\ENDIF
\STATE \textit{\underline{Refinement}}
\STATE Propose $\{\vc{x}_k^*\} \sim q_2\left(\vc{x}_k|\vc{x}_k^{m},\vc{x}_{k-1}^{m}\right)$
\STATE Compute $\psi_2(u,\vc{x}_k^*,\vc{x}_k^{m},\vc{x}_{k-1}^{m})$\\ \hspace{18mm}=$\frac{1}{M_k}\log\left[u\frac{p(\vc{x}_k^{m}|\vc{x}_{k-1}^{m})q_2\left(\vc{x}_k^*|\vc{x}_k^{m},\vc{x}_{k-1}^{m}\right)}{p(\vc{x}_k^*|\vc{x}_{k-1}^m)q_2\left(\vc{x}_k^{m}|\vc{x}_k^{*},\vc{x}_{k-1}^{m}\right)}\right]$
\STATE Compute $\Lambda_2^{M_k}(\vc{x}_k^{m},\vc{x}_k^*) = \frac{1}{M_k}\sum_{i=1}^{M_k}\log\left[\frac{p(\vc{z}_k^i|\vc{x}_k^*)}{p(\vc{z}_k^i|\vc{x}_k^{m})}\right]$
\IF{$\Lambda_2^{M_k}(\vc{x}_k^*,\vc{x}_k^{m}) >\psi_2(u,\vc{x}_k^*,\vc{x}_k^{m},\vc{x}_{k-1}^{m})$}
\STATE $\vc{x}_k^m = \vc{x}_k^*$
\ENDIF
\IF{$m > N_{burn}$}
\STATE $\vc{x}_k^{(m-N_{burn})} = \vc{x}_k^m$
\ENDIF 
\ENDFOR
\ENDFOR
\STATE $\hat{p}(\vc{x}_k|\vc{z}_{1:k}) = \frac{1}{N_p}\sum_{j=1}^{N_p}\delta(\vc{x}_k-\vc{x}_k^{(j)})$
\end{algorithmic}
\end{algorithm}
\subsection{Adaptive Subsampling}
In standard sequential MCMC, it is required to perform $2NM_k$ calculations of the likelihood at each time step. This is highlighted in the computation of the log likelihood ratio, $\Lambda_1^{M_k}(\cdot)$ and $\Lambda_2^{M_k}(\cdot)$, in Algorithm~\ref{alg1}. When $M_k$ is very large, the log likelihood ratio becomes the most computationally expensive step of the algorithm. To reduce the computational complexity, we introduce a Monte Carlo (MC) approximation for the log likelihood ratio:  
\begin{equation}
\Lambda^{S_{m,k}}_1(\vc{x}_k^{m-1},\vc{x}_k^*) = \frac{1}{S_{m,k}}\sum_{i=1}^{S_{m,k}} \log\left[\frac{p(\vc{z}_{k}^{i,*}|\vc{x}_k^*)}{p(\vc{z}_{k}^{i,*}|\vc{x}_k^{m-1})}\right]\label{MCappr}
\end{equation}
where the set $\vc{z}_k^* = \{\vc{z}_{k}^{1,*},...,\vc{z}_{k}^{S_{m,k},*}\}$ is drawn uniformly without replacement from the original set of $M_k$ measurements.

The difficulty which arises is in selecting a minimum value for $S_{m,k}$ that results in a set of subsampled measurements that contain enough information to make the correct decision in the MH step. To overcome this difficulty in standard MCMC for static inference, the authors in \cite{Bardenet2014} proposed to use concentration inequalities which provide a probabilistic bound on how functions of independent random variables deviate from their expectation. In this case, the independent random variables are the log likelihood ratio terms. Thus, it is possible to obtain a bound on the deviation of the MC approximation in \eqref{MCappr} from the complete log likelihood ratio:
\begin{equation}
P(|\Lambda^{S_{m,k}}_1(\vc{x}_k^{m-1},\vc{x}_k^*)-\Lambda^{M_k}_1(\vc{x}_k^{m-1},\vc{x}_k^*)|\! \leq c_{S_{m,k}})\! \geq \! 1-\delta_{S_{m,k}}
\end{equation}
where $\delta_{S_{m,k}} >0$, and $c_{S_{m,k}}$ is dependent on which inequality is used. There are several inequalities which could be used, in this paper we make use of the  empirical Bernstein inequality \cite{audibert2009, Bardenet2015}, which results in:
\begin{equation}
c_{S_{m,k}} = \sqrt{\frac{2V_{S_{m,k}}\log({3/ \delta_{S_{m,k}}})}{{S_{m,k}}}} + \frac{3R_k\log({3/ \delta_{S_{m,k}}})}{{S_{m,k}}} \label{ct}
\end{equation}
where $V_{S_{m,k}}$ represents the sample variance of the log likelihood ratio, and $R_k$ is the range given by
\begin{eqnarray}
\nonumber R_k=\max_{1\le i \le {M_k}}\!\left\{\!\log\!\left[ \frac{p(\vc{z}_{k}^{i}|\vc{x}_k^*)}{p(\vc{z}_{k}^{i}|\vc{x}_k^{m-1})}\right]\!\right\}-\\\min_{1\le i \le {M_k}}\!\left\{\log\! \left[\frac{p(\vc{z}_{k}^{i}|\vc{x}_k^*)}{p(\vc{z}_{k}^{i}|\vc{x}_k^{m-1})}\right]\right\}\!\!\!\!\!\!\!\!\!\!\!\!\!\!\!\!\!\!\!\!\!\!\!\!
\end{eqnarray}

Looking back at the standard sequential MCMC approach, we find that the joint draw is accepted based on the condition $\Lambda_1^{M_k}(\vc{x}_k^*,\vc{x}_k^{m-1})>\psi_1(u,\vc{x}_k^*,\vc{x}_{k-1}^*,\vc{x}_k^{m-1},\vc{x}_{k-1}^{m-1})$. It is required to relate this expression in terms of the MC approximation of \eqref{MCappr}. Since the MC approximation is bounded, we can state that it is not possible to make a decision when the value of $\psi_1(u,\vc{x}_k^*,\vc{x}_{k-1}^*,\vc{x}_k^{m-1},\vc{x}_{k-1}^{m-1})$ falls within the region specified by the bound. Thus it is required that $|\Lambda^{S_{m,k}}_1(\vc{x}_k^{m-1},\vc{x}_k^*)-\psi_1(u,\vc{x}_k^*,\vc{x}_{k-1}^*,\vc{x}_k^{m-1},\vc{x}_{k-1}^{m-1})|>c_{S_{m,k}}$ in order to be able to make a decision, with probability at least $1-\delta_{S_{m,k}}$. 

This forms the underlying principal for the creation of a stopping rule \cite{Bardenet2014,Mnih2008}. Let $\delta \in (0,1)$ be a user specified input parameter. 
The idea is to sequentially increase the size of ${S_{m,k}}$ while at the same time checking if the stopping criterion, $|\Lambda^{S_{m,k}}_1(\vc{x}_k^{m-1},\vc{x}_k^*)-\psi_1(u,\vc{x}_k^*,\vc{x}_{k-1}^*,\vc{x}_k^{m-1},\vc{x}_{k-1}^{m-1})|>c_{S_{m,k}}$, is met. If the stopping criterion is never met, then this will result in ${S_{m,k}} = {M_k}$, i.e requiring the evaluation of all the measurements.  Selecting $\delta_{S_{m,k}} = \frac{p-1}{p{S_{m,k}}^p}\delta$ results in $\sum_{{S_{m,k}}\ge1} \delta_{S_{m,k}} \leq \delta$. The event
\begin{equation}
\mathcal{E} =\!\! \bigcap_{{S_{m,k}}\ge1}\!\!\left\{\!|\Lambda^{S_{m,k}}_1(\vc{x}_k^{m-1},\vc{x}_k^*)-\Lambda^{M_k}_1(\vc{x}_k^{m-1},\vc{x}_k^*)| \leq c_{S_{m,k}}\!\right\}
\end{equation}
thus holds with probability at least $1-\delta$ by a union bound argument. 

This iterative procedure allows for an adaptive size of the number of measurements required to be evaluated. However,  there is cause for concern with the definition of the stopping rule. That is the fact that the range, $R_k$, used in the calculation of \eqref{ct}, is dependent on the log likelihood for all $M_k$ measurements. Calculating this range would thus inherently require at least the same number of calculations as in the standard sequential MCMC approach. In certain applications it may be possible to obtain an expression for the range which is independent of the measurements, however, this is not the case for the current application of interest. In order to overcome the computational complexity of the calculation of the range, and to reduce the sample variance $V_{S_{m,k}}$ in the bound, a control variate has been introduced in \cite{Bardenet2015a}, referred to as a proxy:
\begin{equation}
\wp_i(\vc{x}_k^{m-1},\vc{x}_k^*) \approx \log\left[\frac{p(\vc{z}_{k}^{i}|\vc{x}_k^*)}{p(\vc{z}_{k}^{i}|\vc{x}_k^{m-1})}\right].
\end{equation}
Thus the MC approximation in \eqref{MCappr} is augmented into
\begin{eqnarray}
\!\!\!\!\!\!\!\!\!\Lambda^{S_{m,k}}_1(\vc{x}_k^{m-1},\vc{x}_k^*)\! =\! \frac{1}{S_{m,k}}\sum_{i=1}^{S_{m,k}}\nonumber \log\!\left[\frac{p(\vc{z}_{k}^{i,*}|\vc{x}_k^*)}{p(\vc{z}_{k}^{i,*}|\vc{x}_k^{m-1})}\right]\!\\-\wp_i(\vc{x}_k^{m-1},\vc{x}_k^*).\!\!\!\!\!\!\!\!\!
\end{eqnarray}
It is required to amend the MH acceptance accordingly to take the inclusion of the proxy into account.  

We propose using a first order Taylor series as an approximation for the log likelihood, $\ell_i(\vc{x}) = \log p(\vc{z}^i|\vc{x})$, given as
\begin{equation}
\hat{\ell}_i(\vc{x}) = \ell_i(\vc{x}^+) + (\nabla\ell_i)_{\vc{x}^+}^T\cdot(\vc{x}-\vc{x}^+),
\end{equation}
where $ (\nabla\ell_i)_{\vc{x}^+}$ represents the gradient of ${\ell}_i(\vc{x})$ evaluated at $\vc{x}^+$. This results in the following form of the proxy
\begin{eqnarray}
\nonumber \wp_i(\vc{x}_k^{m-1},\vc{x}_k^*)\!\! \!\! \! &= \hat{\ell}_i(\vc{x}_k^*)-\hat{\ell}_i(\vc{x}_k^{m-1}), \\ \label{proxycalc}
&=  (\nabla\ell_i)_{\vc{x}^+}^T\cdot(\vc{x}_k^* - \vc{x}_k^{m-1}).\!\! \!\! \!\! \!\! \!\! \!\!\!\! \!  
\end{eqnarray}
With the inclusion of the proxy, the range, $R_k$, is now computed as,
\begin{eqnarray}
\nonumber R_k = \max_{1\le i \le {M_k}}\left\{\frac{p(\vc{z}_{k}^{i}|\vc{x}_k^*)}{p(\vc{z}_{k}^{i}|\vc{x}_k^{m-1})}-\wp_i(\vc{x}_k^{m-1},\vc{x}_k^*)\right\}\\
-\min_{1\le i \le {M_k}}\left\{\frac{p(\vc{z}_{k}^{i}|\vc{x}_k^*)}{p(\vc{z}_{k}^{i}|\vc{x}_k^{m-1})}-\wp_i(\vc{x}_k^{m-1},\vc{x}_k^*)\right\}.\!\! \!\!\!\!\! \!\! \!  
\end{eqnarray}
We can derive an upper bound for the range, $R_k^B$, i.e where $R_k^B~\ge~R_k$, which can be computed efficiently 
\begin{align}
R_k^B &= 2\max_{1\le i \le {M_k}}\!\left\{\left|\log\left[ \frac{p(\vc{z}_{k}^{i}|\vc{x}_k^*)}{p(\vc{z}_{k}^{i}|\vc{x}_k^{m-1})}\right]-\wp_i(\vc{x}_k^{m-1},\vc{x}_k^*)\right|\right\}\nonumber \\
&= 2\max_{1\le i \le {M_k}}\!\left\{\left|\ell_i(\vc{x}_k^*)\! -\!\ell_i(\vc{x}_k^{m-1}) \!-\!\hat{\ell}_i(\vc{x}_k^*)\!+\!\hat{\ell}_i(\vc{x}_k^{m-1})  \right|\right\} \nonumber\\
&= 2\max_{1\le i \le {M_k}}\!\left\{\left| B_k({\vc{x}_k^{m-1}}) - B_k({\vc{x}_k^{*}}) \right|\right\} 
\end{align}
where $B_k(\vc{x}) = {\ell}_i(\vc{x}) - \hat{\ell}_i(\vc{x})$ is the remainder of the Taylor approximation. The Taylor-Lagrange inequality  gives us an upper bound on the remainder term. More specifically, if $|f^{(n+1)}(\vc{x})| \le Y$, then $|B_k(\vc{x})| \le \frac{Y|\vc{x}-\vc{x}^+|^{n+1}}{(n+1)!}$, where in our case $n+1 = 2$. Upper bounding the Taylor remainder finally results in the following upper bound on the range
\begin{align}
R_k^B = 2\left|\left|B_k({\vc{x}_k^{m-1}})\right| + \left|B_k({\vc{x}_k^{*}})\right| \right|,\label{UB} 
\end{align}
which is dependent on the maximum of the Hessian of the log likelihood, $Y$.
The complete adaptive subsampling sequential MCMC approach is illustrated by Algorithms~\ref{SS_SMCMC} and \ref{Confidence_Sampler}.
\begin{algorithm}[!ht]
\caption{Adaptive Subsampling Sequential Markov Chain Monte Carlo}
\label{SS_SMCMC}
\begin{algorithmic}[1]
\STATE Initialize particle set: $\{\vc{x}_0^{(j)}\}_{j=1}^{N_p}$
\STATE Determine initial proxy parameters. \label{proxparaminit} (See Section \ref{CCC} for more details.)
\FOR{$k$ = 1,...,$T$}
\FOR{$m$ = 1,...,$N$}
\STATE Update proxy parameters. \label{proxparam} (See Section \ref{CCC} for more details.)
\STATE \textit{\underline{Joint Draw}}
\STATE Propose $\{\vc{x}_k^*,\vc{x}_{k-1}^*\} \sim q_1\left(\vc{x}_k,\vc{x}_{k-1}|\vc{x}_k^{m-1},\vc{x}_{k-1}^{m-1}\right)$
\STATE Compute $\psi_1(u,\vc{x}_k^*,\vc{x}_{k-1}^*,\vc{x}_k^{m-1},\vc{x}_{k-1}^{m-1})$ \\$= \frac{1}{M_k}\log\biggl[u\frac{p(\vc{x}_k^{m-1}|\vc{x}_{k-1}^{m-1})p(\vc{x}_{k-1}^{m-1}|\vc{z}_{1:k-1})}{p(\vc{x}_k^*|\vc{x}_{k-1}^*)p(\vc{x}_{k-1}^*|\vc{z}_{1:k-1})}\times$\\ \hspace{40mm}$\frac{q_1\left(\vc{x}_k^*,\vc{x}_{k-1}^*|\vc{x}_k^{m-1},\vc{x}_{k-1}^{m-1}\right)}{q_1\left(\vc{x}_k^{m-1},\vc{x}_{k-1}^{m-1}|\vc{x}_k^*,\vc{x}_{k-1}^*\right)}\biggr]$

\STATE Compute $\Lambda_1^{S_{m,k}}(\vc{x}_k^*,\vc{x}_k^{m-1})$ and $\{\wp_i(\vc{x}_k^{m-1},\vc{x}_k^*)\}_{i=1}^{M_k}$ with the routine described by Algorithm \ref{Confidence_Sampler}. 
\IF{{$\Lambda_1^{S_{m,k}}(\vc{x}_k^*,\vc{x}_k^{m-1})$ \\ \hspace{15mm}$>\psi_1(u,\vc{x}_k^*,\vc{x}_{k-1}^*,\vc{x}_k^{m-1},\vc{x}_{k-1}^{m-1})$ \\ \hspace{29mm}$-\frac{1}{M_k}\sum_{i=1}^{M_k}\wp_i(\vc{x}_k^{m-1},\vc{x}_k^*)$}}
\STATE $\{\vc{x}_k^m,\vc{x}_{k-1}^m\} = \{\vc{x}_k^*,\vc{x}_{k-1}^*\}$
\ELSE
\STATE $\{\vc{x}_k^m,\vc{x}_{k-1}^m\} = \{\vc{x}_k^{m-1},\vc{x}_{k-1}^{m-1}\}$
\ENDIF
\STATE \textit{\underline{Refinement}}
\STATE Propose $\{\vc{x}_k^*\} \sim q_2\left(\vc{x}_k|\vc{x}_k^{m},\vc{x}_{k-1}^{m}\right)$
\STATE Compute $\psi_2(u,\vc{x}_k^*,\vc{x}_k^{m},\vc{x}_{k-1}^{m})$\\ \hspace{20mm}=$\frac{1}{M_k}\log\left[u\frac{p(\vc{x}_k^{m}|\vc{x}_{k-1}^{m})q_2\left(\vc{x}_k^*|\vc{x}_k^{m},\vc{x}_{k-1}^{m}\right)}{p(\vc{x}_k^*|\vc{x}_{k-1}^m)q_2\left(\vc{x}_k^{m}|\vc{x}_k^{*},\vc{x}_{k-1}^{m}\right)}\right]$
\STATE Compute $\Lambda_2^{S_{m,k}}(\vc{x}_k^{m},\vc{x}_k^*)$ and $\{\wp_i(\vc{x}_k^{m},\vc{x}_k^*)\}_{i=1}^{M_k}$ with the routine described by Algorithm \ref{Confidence_Sampler}.
\IF{$\Lambda_2^{S_{m,k}}(\vc{x}_k^*,\vc{x}_k^{m}) >\psi_2(u,\vc{x}_k^*,\vc{x}_k^{m},\vc{x}_{k-1}^{m})$\\ \hspace{27mm}$-\frac{1}{M_k}\sum_{i=1}^{M_k}\wp_i(\vc{x}_k^{m},\vc{x}_k^*)$}
\STATE $\{\vc{x}_k^m\} = \{\vc{x}_k^*\}$
\ENDIF
\IF{$m > N_{burn}$}
\STATE $\vc{x}_k^{(m-N_{burn})} = \vc{x}_k^m$
\ENDIF 
\ENDFOR
\ENDFOR
\STATE $\hat{p}(\vc{x}_k|\vc{z}_{1:k}) = \frac{1}{N_p}\sum_{j=1}^{N_p}\delta(\vc{x}_k-\vc{x}_k^{(j)})$
\end{algorithmic}
\end{algorithm}

\begin{algorithm}[!ht]
\caption{Adaptive Subsampling Routine}
\label{Confidence_Sampler}
\begin{algorithmic}[1]
\STATE Given: The current and proposed states of the Markov chain, $\{\vc{x}_k$, $\vc{x}_k^*\}$, the complete measurement set, $\vc{z}_k = \{\vc{z}_k^1,...,\vc{z}_k^{M_k}\}$, $\delta$, and $\psi(\cdot)$. 
\STATE Initialise: Number of sub-sampled measurements, $S_{m,k}~=~0$, Approximate log likelihood ratio subtracted by proxy, $\Lambda~=~0$, set of sub-sampled measurements, $\vc{z}_k^*~=~\emptyset$, initial batchsize, $b~=~1$, while loop counter, $w~=~0$.
\STATE Compute an upper bound for the range, $R_k^B$, according to~\eqref{UB}.
\STATE Compute the proxy, $\{\wp_i(\vc{x}_k,\vc{x}_k^*)\}_{i=1}^{M_k}$, according to \eqref{proxycalc}.
\STATE DONE = FALSE
\WHILE{DONE == FALSE}
\STATE $w = w + 1$
\STATE $\{\vc{z}_k^{S_{m,k}+1,*},...,\vc{z}_k^{b,*}\} \sim_{w/repl.}\vc{z}_k \setminus \vc{z}_k^*$
\STATE $\vc{z}_k^* = \vc{z}_k^* \cup \{\vc{z}_k^{S_{m,k}+1,*},...,\vc{z}_k^{b,*}\}$
\STATE $\Lambda\!\! =\! \!\frac{1}{b}\!\!\left(\!\!S_{m,k}\Lambda\! +\! \!\sum_{i=S_{m,k} + 1}^b\!\! \left[\log \! \frac{p(\vc{z}_{k}^{i,*}\!|\vc{x}_k^*)}{p(\vc{z}_{k}^{i,*}\!|\vc{x}_k)}\!-\!\wp_i(\vc{x}_k,\vc{x}_k^*)\!\right]\!\right)$
\STATE $S_{m,k} = b$
\STATE $\delta_{w} = \frac{p-1}{p{w}^p}\delta$
\STATE Compute $c$ according to \eqref{ct} utilising $\delta_w$.
\STATE $b = \gamma S_{m,k}\wedge M_k$
\IF{$|\Lambda + \frac{1}{M_k}\sum_{i=1}^{M_k}\wp_i(\vc{x}_k,\vc{x}_k^*) - \psi(\cdot)|\ge c$ \OR $S_{m,k} == M_k$}
\STATE DONE = TRUE
\ENDIF
\ENDWHILE
\RETURN $\Lambda$ and $\{\wp_i(\vc{x}_k,\vc{x}_k^*)\}_{i=1}^{M_k}$
\end{algorithmic}
\end{algorithm}

\section{Application to Target Tracking in Complex Systems}
\subsection{Target and Sensor Modelling}
In this application the state vector consists of the position and velocity of the target in a two dimensional space, $\vc{x}_k~=~[x_k,y_k,\dot{x}_k, \dot{y}_k]^T$.The target motion prediction is performed according to the near constant velocity model. This results in the state transition density having the form
\begin{equation}
p(\vc{x}_k|\vc{x}_{k-1})=\mathcal{N}(\vc{x}_k|\vc{A}_k\vc{x}_{k-1},\vc{Q}_k),
\end{equation}
where $\mathcal{N}(\cdot)$ represents the normal distribution, and matrices $\vc{A}_k$ and $\vc{Q}_k$ are defined as $\vc{A}_k = \left[
                                                           \begin{array}{cc}
                                                             \vc{I}_2 & T_s\vc{I}_2 \\
                                                             \vc{0}_2 & \vc{I}_2 \\
                                                           \end{array}
                                                         \right]$ and $\vc{Q}_k~=~\sigma^2_x\left[
                                                           \begin{array}{cc}
                                                             (T_s^3/3)\vc{I}_2 & (T_s^2/2)\vc{I}_2 \\
                                                             (T_s^2/2)\vc{I}_2 & T_s\vc{I}_2 \\
                                                           \end{array}
                                                         \right]$, where $T_s = t_k - t_{k-1}$.
                                                          
In this application, the total number of measurements received is given by $M_k = M_k^x + M_k^c$, where $M_k^x$ represents the number of target measurements, and $M_k^c$ represents the number of clutter measurements. The number of target and clutter measurements are Poisson distributed with mean $\lambda_X$ and $\lambda_C$ respectively. The likelihood density thus takes the form~\cite{Gilholm2005}:
\begin{equation}
p(\vc{z}_k|\vc{x}_k)\propto \prod_{i=1}^{M_k} \lambda_Xp_X(\vc{z}_k^i|\vc{x}_k) + \lambda_Cp_C(\vc{z}_k^i),\label{genlik}
\end{equation}
where $p_X(\cdot)$ and $p_C(\cdot)$ represent the likelihood of the target and clutter measurements respectively. Each individual measurement represents a point in the two dimensional observation space, $\vc{z}_k^i~=~\left[z_{x,k}^i,z_{y,k}^i\right]^T$. In the case of a measurement from the target, the likelihood is modelled as $p_X(\vc{z}_k^i|\vc{x}_k) = \mathcal{N}(\vc{z}_k^i;\vc{x}_k,\Sigma)$. The clutter measurements are independent of the state of the target and are uniformly distributed in the visible region of the sensor, resulting in the clutter likelihood taking the form of $p_C(\vc{z}_k^i)= \textit{U}_{R_x}(z_{x,k}^i) \textit{U}_{R_y}(z_{y,k}^i)$.

The Taylor approximations used by the proxy in \eqref{proxycalc} are dependent on the gradient and Hessian of the log likelihood for individual measurements. Substituting the terms for the target and clutter likelihood in \eqref{genlik} and taking the logarithm results in the log likelihood for each measurement having the form
\begin{equation}
\ell_i(\vc{x}_k)=\log\left(\lambda_X \mathcal{N}(\vc{z_k^i};\vc{x}_k,\vc{\Sigma}) + \frac{\lambda_C}{A_C}\right),
\end{equation}
where $A_C = R_x \times R_y$ represents the clutter area.
The gradient can then be computed as 
\begin{align}
\nabla\ell_i =\frac{\lambda_X \vc{\Sigma}^{-1}(\vc{z}^i_k-\vc{x}_k)\mathcal{N}(\vc{z}^i_k;\vc{x}_k,\vc{\Sigma})}{\lambda_X \mathcal{N}(\vc{z}^i_k;\vc{x}_k,\vc{\Sigma}) + \frac{\lambda_C}{A_C}}, \label{gradient}
 \end{align}
 and the Hessian is given by
\begin{align}
\vc{H} =&\frac{-\lambda_X \vc{\Sigma}^{-1}\mathcal{N}(\vc{z}_k^i;\vc{x}_k,\vc{\Sigma})\left(\lambda_X \mathcal{N}(\vc{z}_k^i;\vc{x}_k,\vc{\Sigma}) + \frac{\lambda_C}{A_C}\right)}{\left(\lambda_X \mathcal{N}(\vc{z}_k^i;\vc{x}_k,\vc{\Sigma}) + \frac{\lambda_C}{A_C}\right)^2}+\nonumber \\  
&\frac{\frac{\lambda_C \lambda_X}{A_C} \vc{\Sigma}^{-1}(\vc{z}_k^i-\vc{x}_k)\left(\vc{\Sigma}^{-1}(\vc{z}_k^i-\vc{x}_k)\right)^T\mathcal{N}(\vc{z}_k^i;\vc{x}_k,\vc{\Sigma})}{\left(\lambda_X \mathcal{N}(\vc{z}_k^i;\vc{x}_k,\vc{\Sigma}) + \frac{\lambda_C}{A_C}\right)^2} \label{Hessian}
\end{align}
\subsection{Implementation Considerations}
\label{CCC}
The primary difference between the standard and adaptive subsampling sequential MCMC is that the latter requires less evaluations of the log likelihood. However, there are also additional computations which are introduced to achieve this. These calculations are minimal and typically performed for a fraction of the time spent on the calculation of the likelihood, when $M_k$ is sufficiently large, and are thus considered negligible. In this section we discuss these computations in more detail. 

The proxy, given in \eqref{proxycalc} is extremely efficient to compute in comparison to the log likelihood. This is conditioned on the availability of the gradient of the log likelihood (i.e. \eqref{gradient}) evaluated at a specific point.  Currently, we only update this twice per time step  (represented by line \ref{proxparam} in Algorithm \ref{SS_SMCMC}). Once, at the beginning of a time step, where the specific point used is the predicted mean of the Markov chain at the previous time step. Secondly, the current state of the Markov chain after the burn in period. As the number of MCMC particles, $N$, is typically several magnitudes larger than 2, these calculations are considered negligible.

The calculation of an upper bound on the range in \eqref{UB} is also extremely efficient to compute in comparison to the log likelihood. This is conditioned on the availability of the maximum of the Hessian in \eqref{Hessian}. In our application we found that the maximum of the Hessian is independent of the measurements and can hence be determined prior to the running of the algorithm (represented by line \ref{proxparaminit} in Algorithm~\ref{SS_SMCMC}). 

The proposal distribution used for the joint draw step in the tracking scenario is defined as:
\begin{equation}
\small{q_1\left(\vc{x}_k,\vc{x}_{k-1}|\vc{x}_k^{m},\vc{x}_{k-1}^{m}\right)=p(\vc{x}_k|\vc{x}_{k-1})\frac{1}{N_p}\sum_{j=1}^{N_p}\delta(\vc{x}_{k-1}-\vc{x}_{k-1}^{(j)})}.
\end{equation}
The proposal distribution used for the refinement step in the tracking scenario is defined as:
\begin{equation}
\small{q_2\left(\vc{x}_k|\vc{x}_k^{m},\vc{x}_{k-1}^{m}\right)= \mathcal{N}(\vc{x}_k^m,\vc{\Sigma}_q)}.
\end{equation}
The refinement step represents a local move.
\section{Results}
Consider the scenario of a target moving through a highly cluttered environment. A sensor monitoring the target returns multiple target and clutter measurements at each time step. We applied the standard and adaptive subsampling sequential MCMC algorithms for the inference of the latent states of the target over several experiments with different parameters. 

Two different metrics are used to compare the performance of the algorithms. Firstly, the root mean square error (RMSE) of the position. The RMSE for each time step is calculated over a number of independent simulation runs according to
\begin{equation}
RMSE = \sqrt{\frac{1}{N_{I}}\sum_{i=1}^{N_{I}}(\hat{X}_i-X_i)^2},
\end{equation}
where $X_i$ represents the ground truth, $\hat{X}_i$ represents the algorithm estimate, which corresponds to the mean of the $N$ MCMC samples in this application, and $N_{I}$ represents the number of independent runs. The RMSE of the states corresponding to the position are averaged to obtain a single result. The RMSE of the position illustrates the tracking accuracy of the two algorithms. 

The second metric is the normalized number of sub-sampled measurements required for likelihood calculations. 
\begin{equation}
D = \frac{1}{T}\sum_{k=1}^T\frac{ \sum_{m=1}^N(S_{m,k})_{JD} + (S_{m,k})_{R}}{2NM_k}
\end{equation}
where $(S_{m,k})_{JD}$ and $(S_{m,k})_{R}$ refer to the number of sub-sampled measurements from the joint draw step and refinement step respectively.  The standard sequential MCMC algorithm requires to evaluate the likelihood $2NM_k$ times at each time step, this corresponds to $D = 1$. Thus the $D$ value is only shown for the adaptive subsampling sequential MCMC algorithm. It illustrates the fraction of likelihood evaluations which are required at each time step versus the standard sequential MCMC algorithm. 
\subsection{Parameters}
The following parameters, unless otherwise specified, were used for all experiments. Simulation parameters: $N = 500$, $N_{burn} = 125$, $T = 20$, $N_I = 50$, $\vc{\Sigma}_q = 0.01\vc{I}$. Motion model parameters: $T_s = 1$, $\sigma_x = 0.5$. Target observation model  parameters: $\lambda_X = 500$, $\vc{\Sigma} = \vc{I}$. Clutter parameters: $\lambda_C = 2000$, $A_c = 4\times10^4$. Subsampling parameters: $\gamma = 1.2$, $\delta = 0.1$, $p = 2$. 
\subsection{Performance Evaluation}
The first experiment illustrates the performance of the algorithms for different values of the mean total number of measurements in Fig. 1. The ratio between the mean number of clutter measurements and mean number of target measurements is fixed at 4:1. The RMSEs of the algorithms are in agreement, however, it is noted that an increase in the total mean number of measurements results in substantial computational savings. The amount of computational saving is as high as 80\% with no significant loss in tracking performance.

\begin{figure}[h]
\centering
\subfloat[RMSE comparison for different values of mean number of total measurements. The dotted lines represent the results from the standard sequential MCMC, and the full lines represent the results from the adaptive subsampling sequential MCMC.]{
\includegraphics[width = 85mm]{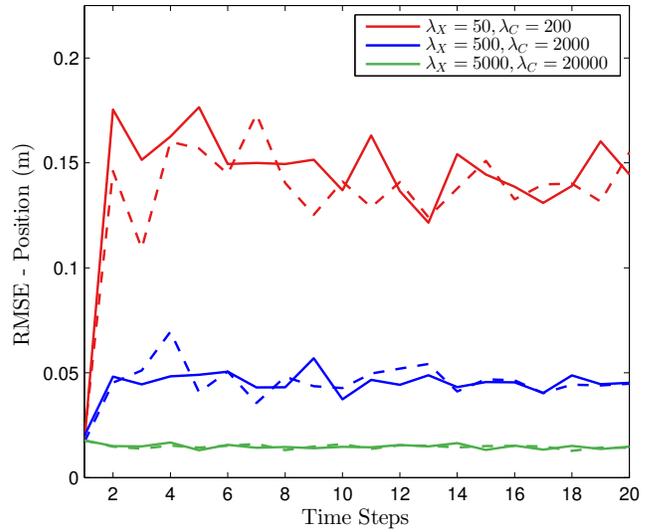}}
\qquad
\subfloat[Comparison of the normalized number of subsampled measurements evaluated in the adaptive subsampling sequential MCMC for different values of mean number of total measurements.]{
\includegraphics[width = 85mm]{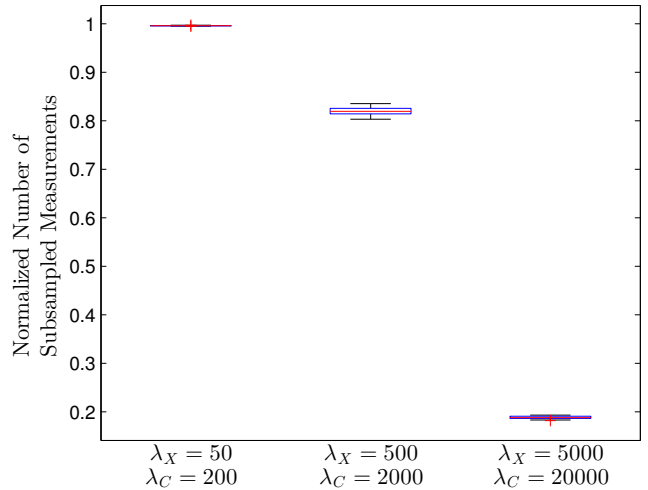}}
\qquad
\caption{Performance comparison for a different mean number of total measurements with a constant clutter to target measurement ratio of 4:1.}
\end{figure}
In Fig. 2 the ratio between the mean number of clutter measurements and mean number of target measurements is varied. This allows for the observation of the performance when there is a varied amount of information about the target present in the measurements. The RMSE results show agreement between the two algorithms with an increase in computational savings when the mean number of target measurements is higher. 
\begin{figure}[h]
\centering
\subfloat[RMSE comparison of different values of mean target measurements. The dotted lines represent the results from the standard sequential MCMC, and the full lines represent the results from the adaptive subsampling sequential MCMC.]{
\includegraphics[width = 85mm]{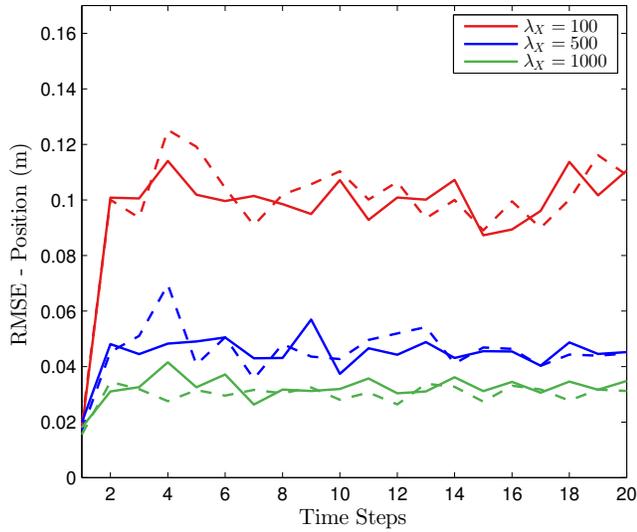}}
\qquad
\subfloat[Comparison of the normalized number of subsampled measurements evaluated in the adaptive subsampling sequential MCMC for different values of mean target measurements.]{
\includegraphics[width = 85mm]{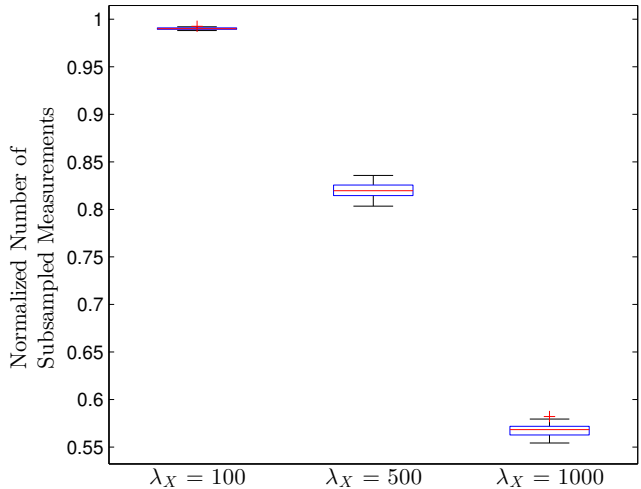}}
\caption{Performance comparison for a different number of mean clutter to target measurements ratios.}
\end{figure}

Fig. 3 illustrates the influence of varying the covariance matrix of the target observation model. The RMSEs of the two algorithms are in agreement. It is noted that a smaller computational saving is incurred as the measurement model becomes more precise. This result seems counter-intuitive. The reason for this is due to the Taylor approximation for the proxy. The upper bound for the range, $R^B_k$, becomes a weaker bound as the observation model becomes more peaked.
\begin{figure}[h]
\centering
\subfloat[RMSE comparison for different measurement covariance matrices. The dotted lines represent the results from the standard sequential MCMC, and the full lines represent the results from the adaptive subsampling sequential MCMC.]{
\includegraphics[width = 85mm]{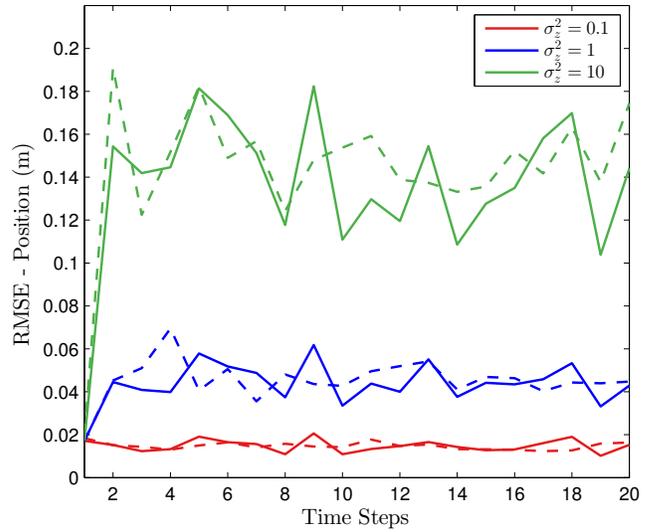}}
\qquad
\subfloat[Comparison of the normalized number of subsampled measurements evaluated in the adaptive subsampling sequential MCMC for different covariance matrices.]{
\includegraphics[width = 85mm]{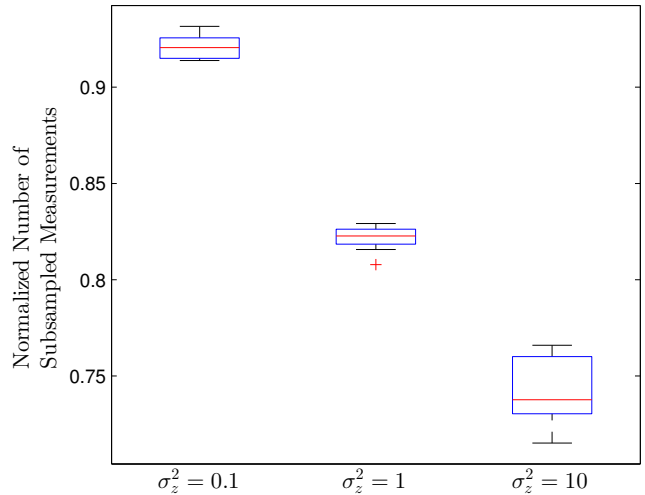}}
\caption{Performance comparison for different covariance matrices where $\vc{\Sigma} = \sigma_z^2 \vc{I}$.}
\end{figure}
\section{Conclusion}
In this paper, we presented an adaptive subsampling sequential MCMC algorithm for target tracking. We have shown that this approach results in substantial computational savings when there is a large number of measurements, and most importantly, without sacrificing tracking performance. 

There is a wide scope for future work. From an application perspective, considering a multi-target scenario with different levels of clutter, and also from an algorithmic perspective, further research on the influence and implementation of a more efficient proxy.
\section*{Acknowledgments}
We would like to thank R\'{e}mi Bardenet for the constructive discussions on this work. We also acknowledge the support from the UK Engineering and Physical Sciences Research Council (EPSRC) via the Bayesian Tracking and Reasoning over Time (BTaRoT) grant EP/K021516/1 and EC Seventh Framework Programme [FP7 2013-2017] TRAcking in compleX sensor systems (TRAX) Grant agreement no.: 607400.

\bibliographystyle{IEEEtran}
\bibliography{Allan}



%

\end{document}